\documentclass[aps,twocolumn,9pt,a4paper,superscriptaddress]{revtex4-2}

\usepackage[charter]{mathdesign}
\DeclareSymbolFont{usualmathcal}{OMS}{cmsy}{m}{n}
\DeclareSymbolFontAlphabet{\mathcal}{usualmathcal}
\usepackage{amsmath}
\usepackage[pdftex]{graphicx}
\usepackage{microtype}
\usepackage{dsfont}
\usepackage{bm}
\usepackage[svgnames]{xcolor}
\usepackage[colorlinks=True,linkcolor=DarkRed,citecolor=Orchid,urlcolor=MediumBlue,
	pdfstartview=FitH,bookmarks=False,pdfpagemode=UseNone]{hyperref}

\counterwithout{equation}{section}

\newcommand{\vect}[1]{\bm{#1}}
\newcommand{\panel}[1]{{\fontfamily{phv}\selectfont\textbf{#1}}}
\def\grad{\nabla}

\def\itop#1{\vtop{\null\hbox{#1}}}

\begin{document}
%TC:ignore
\title{Black-hole spectroscopy from a giant quantum vortex}

\author{Pietro Smaniotto}
\affiliation{School of Mathematical Sciences, University of Nottingham, University Park, Nottingham, NG7 2RD, UK}
\affiliation{Nottingham Centre of Gravity, University of Nottingham,
University Park, Nottingham NG7 2RD, UK}

\author{Leonardo Solidoro}
\affiliation{School of Mathematical Sciences, University of Nottingham, University Park, Nottingham, NG7 2RD, UK}
\affiliation{Nottingham Centre of Gravity, University of Nottingham,
University Park, Nottingham NG7 2RD, UK}

\author{Patrik \v{S}van\v{c}ara}
\affiliation{School of Mathematical Sciences, University of Nottingham, University Park, Nottingham, NG7 2RD, UK}
\affiliation{Nottingham Centre of Gravity, University of Nottingham,
University Park, Nottingham NG7 2RD, UK}
\affiliation{Institut N\'{e}el, CNRS/UGA, 25 avenue des Martyrs, 38042 Grenoble, France}

\author{Sam~Patrick}
\affiliation{Department of Physics and Astronomy, University of Manchester, Manchester M13 9PL, UK}
\affiliation{Photon Science Institute, Alan Turing Building, University of Manchester, Manchester M13 9PY, UK}
\affiliation{Department of Physics, King’s College London, University of London, Strand, London, WC2R 2LS, UK}

\author{Maur\'icio Richartz}
\affiliation{Centro de Matem\'atica, Computa\c c\~ao e Cogni\c c\~ao,
Universidade Federal do ABC (UFABC), 09210-170 Santo Andr\'e, S\~ao Paulo, Brazil}

\author{Carlo F. Barenghi}
\affiliation{School of Mathematics, Statistics and Physics, Newcastle University, Newcastle upon Tyne, NE1 7RU, UK}

\author{Ruth Gregory}
\affiliation{Department of Physics, King’s College London, University of London, Strand, London, WC2R 2LS, UK}
\affiliation{Perimeter Institute, 31 Caroline Street North, Waterloo, ON, N2L 2Y5, Canada}

\author{Silke Weinfurtner}
\email{silke.weinfurtner@manchester.ac.uk}
\affiliation{School of Mathematical Sciences, University of Nottingham, University Park, Nottingham, NG7 2RD, UK}
\affiliation{Nottingham Centre of Gravity, University of Nottingham,
University Park, Nottingham NG7 2RD, UK}
\affiliation{Department of Physics and Astronomy, University of Manchester, Manchester M13 9PL, UK}
\affiliation{Photon Science Institute, Alan Turing Building, University of Manchester, Manchester M13 9PY, UK}
\affiliation{Perimeter Institute, 31 Caroline Street North, Waterloo, ON, N2L 2Y5, Canada}
\affiliation{Centre for the Mathematics and Theoretical Physics of Quantum Non-Equilibrium Systems, University of Nottingham, Nottingham, NG7 2RD, UK}

\maketitle
\onecolumngrid
\vspace{-0.8cm}

\par\noindent
Black-hole spectroscopy~\cite{bertispectroscopy} aims to infer the fundamental properties of black holes by analysing the spectrum of gravitational waves emitted as they settle into equilibrium. These resonances, known as quasinormal modes (QNMs), decay rapidly, which limits the time-domain analysis of gravitational-wave data or numerical simulations to the longest-lived mode~\cite{Nee:2023osy}, except for a particularly loud event~\cite{abac2025}. Owing to the analogy between fields in curved spacetime and waves propagating in a flowing medium, QNMs can be equally excited in a laboratory~\cite{barcelo2011}. In these finite-sized systems, the QNM spectrum is expected to alter~\cite{solidoro2024}: compared to their counterparts in unbounded settings, the real frequencies of QNMs shift while their damping rates (imaginary frequencies) reduce, thereby enhancing their detectability. Here we show that multiple QNMs can be extracted from noise-driven interface waves surrounding a giant quantum vortex in superfluid \mbox{helium-4} \cite{svancara2024}, which emulates a spacetime geometry indicative of a rotating black hole~\cite{visser2005}. By resolving waves with different azimuthal periodicity, we find that both fundamental modes and their higher-frequency overtones are excited, and oscillate at frequencies given by the size of our system. Since similar effects may arise in astrophysical scenarios due to the interstellar medium or dark matter~\cite{Barausse:2014tra, cardoso2022}, gravity simulators now complement numerical and observational approaches to black-hole spectroscopy.
\bigskip

\twocolumngrid
%TC:endignore

\noindent
The broad range of fluid systems simulating curved spacetime geometries developed to date, both classical \cite{weinfurtner2011,euve2016,torres2017} and quantum \cite{hung2013,Eckel2018,jacquet2020,navon2021,viermann2022}, has enabled experimental exploration of the rich phenomenology of black-hole physics. Phenomena including Hawking radiation  \cite{weinfurtner2011,euve2016,steinhauer2021} and rotational superradiance \cite{torres2017,cromb2020,braidotti2022} have been directly observed in hydrodynamic experiments. The emission of waves related to QNMs has been studied numerically in ideal fluids \cite{Cardoso:2004fi,patrick2018,Torres:2019sbr}, Bose–Einstein condensates \cite{geelmuyden2022}, quantum fluids of light \cite{solnyshkov2019,jacquet2023,burkhard2025stimulated}, and optical solitons \cite{burgess2024}. In these models, the boundary conditions are open, permitting energy and information to escape to infinity. Experimental observations of QNMs in a water tank \cite{torres2020} likewise result in an effectively open system due to linear dissipation and the geometry of the boundary.

Here we examine the occurrence of QNMs in a newly developed gravity simulator \cite{svancara2024} that employs superfluid helium-4, a quantum liquid. The underlying rotating curved spacetime is induced by a giant quantum vortex that is stabilised by draining the superfluid through an opening in the centre of a cylindrical container (see Methods for details). Mechanical vibrations of the set-up induce interface noise, i.e. broadband fluctuations of the superfluid-vapour interface composed of many superimposed waves. We capture these fluctuations and probe the effective spacetime by implementing a non-invasive, space- and time-resolved readout technique \cite{moisy2009,wildeman2018}. Moreover, the incredibly weak damping of superfluid waves, the relatively small size of our system, and the ability to read out interface fluctuations over an extended period of time, mean that boundary effects cannot be ignored. Using a mathematical toy-model capturing relevant features of the spatially confined black hole spacetime~\cite{solidoro2024}, it has been shown that QNM signatures persist in the simulated spectrum, with their frequencies shifting along the real axis while their damping rates decrease but do not completely vanish. This reduction in damping enhances the prospect of detecting multiple QNMs simultaneously, including the longest-lived fundamental mode and its overtones.

Although our earlier work suggested the excitation of a QNM in superfluid helium \cite[Fig.~5c]{svancara2024}, this identification remained tentative, limited by both imaging resolution and noise control, while the theoretical framework lacked the precision required for a quantitative comparison with the observed mode frequencies. By reducing narrow-band noise injections and improving imaging sensitivity (see Methods), we obtained vortex configurations that display multiple QNMs. In parallel, we extended the modelling framework to encompass interactions between surface waves, the giant quantum vortex and the outer boundary, allowing quantitative comparisons between measured excitations and theoretical QNM predictions. This enables us to extend black-hole spectroscopy to the quantitative study of confinement effects on the QNM spectrum.

%TC:ignore
\begin{figure}[htbp!]
    \centering
    \includegraphics[width=\columnwidth]{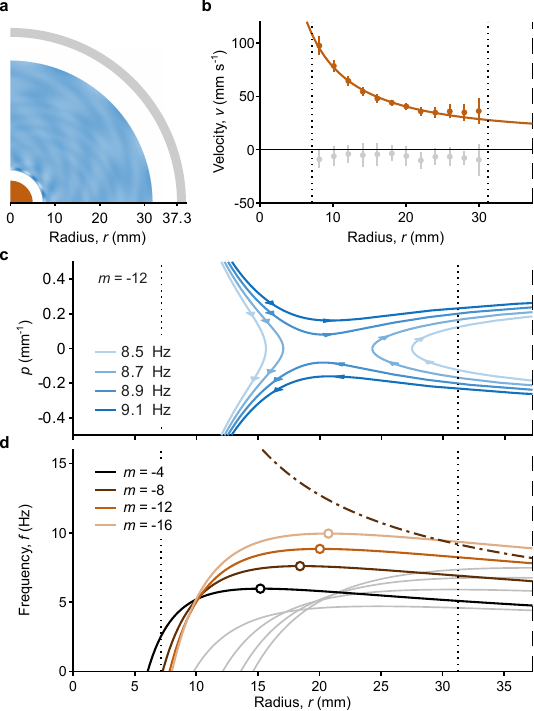}
    \caption{{\bf Flow-induced scattering potential.} \panel{a}~Top view of the experimental zone (one quadrant shown). The orange region represents the drain ($5.0$~mm radius). The blue area, spanning from $7.2$ to $31.2$~mm (black dotted lines in later panels), marks the detection range for interface waves and displays an illustrative snapshot of the superfluid interface. The grey region denotes the outer boundary at $37.3$~mm (black dashed line in later panels). \panel{b}~Velocity components of the superfluid flow as a function of radius. Orange (grey) points represent azimuthal (radial) velocity values averaged over $2.0$-mm radial intervals. The orange line fits Eq.~\eqref{eq:velocity}. Vertical error bars indicate one standard deviation. \panel{c}~Phase-space diagram for the $m = -12$ azimuthal number. The radial component of the wave vector $\vect{k}$ is denoted by $p$. Waves with frequencies below $8.83$~Hz reflect on both sides of the flow-induced effective scattering potential plotted in panel~\panel{d}, while higher-frequency waves are allowed to propagate between the vortex and the outer boundary. \panel{d}~The effective scattering potential (see main text) for four counterrotating azimuthal modes (see legend) and for $m = 8$ (brown dot-dashed line). Local maxima of the potentials (open circles) correspond to light-ring frequencies. If we were to consider a vortex with circulation $2C$ (solid grey lines), the cavity formed between the light ring and the outer boundary would have reduced in size and depth, or would not have formed at all. Conversely, when $C$ is too small, the maximum of the effective potential is deep in the vortex core where the curved spacetime analogy breaks down.}
    \label{fig:intro}
\end{figure}
%TC:endignore

Fig.~\ref{fig:intro}a outlines the geometry of our experiment. The giant quantum vortex is located above the central opening (orange region) and manifests itself by a visible depression of the superfluid interface. Fluctuations of the interface height are resolved within an annular region (blue) that captures approximately $2/3$ of the experimental zone enclosed by a glass wall (grey). Cylindrical symmetry and stationarity of the flow ensure the conservation of frequency $f$ and azimuthal number $m$ of surface waves, with the latter quantifying the wave's periodicity along a circular path. We decompose the interface noise into individual modes labelled by $m$ and $f$ (see Methods) and, using the technique established in \cite{svancara2024}, we exploit the characteristic shape of the noise spectra to determine the superfluid velocity field at the interface, shown in Fig.~\ref{fig:intro}b. The radial velocity component (grey points) is negligible in the fluid layer probed by the waves \cite{andersen2003}, while the azimuthal component (orange points) exhibits a distinct radial dependence. The overall flow field can be written as
\begin{align}
    \vect{v}(r) = \left(\frac{C}{r}+\Omega r\right) \hat{\vect{\theta}}\,,
    \label{eq:velocity}
\end{align}
with the best fit shown in Fig.~\ref{fig:intro}b by an orange line. The dominant term $C/r$ represents an irrotational (vorticity-free) vortex with circulation $C$ located at the origin, while the solid-body rotation term $\Omega r$ contributes to Eq.~\eqref{eq:velocity} less than $13\%$ within our field of view (see Methods). Since circulation in superfluid helium is fundamentally discrete, the core of the vortex consists of a dense cluster of approximately 48,500 singly-quantised vortices, nanometre-thin topological defects within the superfluid \cite{svancara2024}. For the length scales of interest, ranging from $10^{-6}$ to $10^{-2}~\mathrm{m}$, these vortices can be treated as a coarse grained continuum described by Eq.~\eqref{eq:velocity}, with the majority of vortices concentrated above the drain (orange area in Fig.~\ref{fig:intro}a).

Low-frequency waves perceive the presence of this flow field through the effective metric in one time-like and two space-like dimensions,
\begin{equation}
      g_{ij} \propto \begin{pmatrix}-c^2+v^2 & -\bm{v} \\ -\bm{v} & \mathds{1}_{2\times 2}\end{pmatrix},
      \label{eq:metric}
\end{equation}
where $c$ represents the wave propagation speed. Although this metric does not account how dispersion affects high-frequency waves, previous studies have shown that curved-spacetime phenomenology persists in such regimes~\cite{patrick2020}. While the lack of a radial flow in \eqref{eq:velocity} means the horizon is absent from our simulated spacetime, a vortex which rotates sufficiently rapidly possesses the same kinematic features responsible for QNMs of real black holes, as we outline below.

To analyse the wave dynamics, we employ the Wentzel–Kramers–Brillouin (WKB) approximation \cite{buhler2014}, which models interface fluctuations as a superposition of plane waves with a spatially varying wave vector $\vect{k} = p(r) \,\hat{\vect{r}} + (m/r) \,\hat{\vect{\theta}}$ and fixed angular frequency $\omega = 2\pi f$. Without loss of generality, we focus on $\omega>0$ so that $m>0$ ($m<0$) represents waves which corotate (counterrotate) with the vortex.
The waves follow the dispersion relation~\cite{patrick2020},
\begin{align}
    \label{eq:disp}
    \left(\omega-\vect{v}\cdot\vect{k}\right)^2 = F(\vect{k})\,,
\end{align}
where $F(\vect{k})$ is the function that characterises dispersive properties of the medium (see Methods). For given $\omega$ and $m$, we solve Eq.~\eqref{eq:disp} to obtain the radial wavenumber $p$ as a function of radius. Waves propagate in regions where $p$ is a real number, corresponding to oscillatory solutions that travel through the system. In Fig.~\ref{fig:intro}c, we plot $p$ as a function of radius for the fixed counterrotating mode $m = -12$. We identify two distinct behaviours for the flow parameters considered here. At high frequencies, we find a pair of radially ingoing ($p<0$) and outgoing ($p>0$) waves (dark blue lines) that is allowed to propagate between the boundary and the centre of the experimental zone. However, at lower frequencies (light blue lines), the waves starting near the outer boundary (edge of the experimental zone) and near the vortex are reflected back. This behaviour is consistent with the concept of an effective potential barrier that separates regions admitting real and imaginary (evanescent) solutions, respectively. We identify this potential by recasting Eq.~\eqref{eq:disp} as $\omega=\omega_D^\pm$, where
\begin{align}
    \omega_D^\pm(p,m,r) = \underbrace{\frac{mC}{r^2} + m\Omega}_{\vect{v}\cdot\vect{k}}\, \pm \sqrt{F(p,m,r)}\,,
    \label{eq:branch}
\end{align}
and noting that, for $p=0$, Eq.~\eqref{eq:branch} exhibits a local minimum for $\omega_D^+$ and a maximum for $\omega_D^-$. The frequency range between these extrema only contains evanescent solutions, corresponding to imaginary $p$ and exponentially decaying, non-propagating modes. Therefore, the curve $\omega_D^+(p=0,m,r)$ separates the propagating and evanescent regions and acts as this potential barrier (see Methods for a detailed discussion).

We plot the effective potential as a function of radius by solid coloured lines in Fig.~\ref{fig:intro}d. The dot-dashed line, corresponding to the potential of a corotating azimuthal number $m=8$, diverges at small radii, illustrating that corotating modes cannot propagate to the centre of the vortex; in this sense, the vortex core effectively acts as a hard wall. QNMs arise in scenarios where the effective potential features a local maximum \cite{bertispectroscopy}, which occurs only for counterrotating waves in our set-up. The open circles indicate the location of the maximum of the effective potential, which is analogous to the light ring in black hole physics. It marks the unstable boundary between waves that fall into the black hole horizon and those that escape to infinity, and is responsible for the shape of the accretion disk in images of real black holes~\cite{eventhorizon}. In our case, waves near the light ring represent the lowest-energy excitations able to radiate energy from the vortex, and approximate the real part of the fundamental QNM frequency when the WKB approximation holds \cite{cardoso2009}. The presence of a light ring within the system places a strong constraint on the values of $C$. In order to uniquely identify QNMs, we tune the background flow to place the light ring within the detection range.

\begin{figure}
    \centering
    \includegraphics[width=\columnwidth]{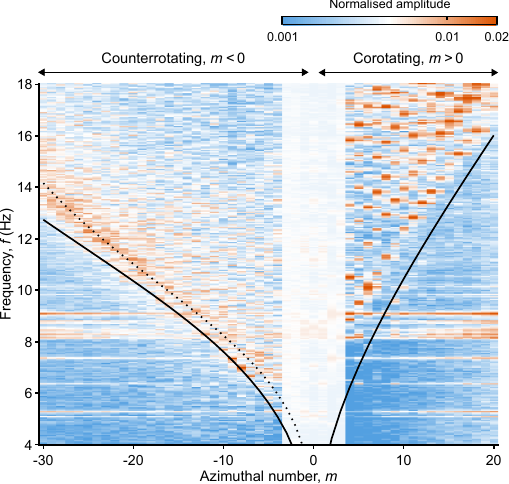}
    \caption{{\bf Landscape of surface excitations.} The spectra of counterrotating ($m<0$) and corotating ($m>0$) waves (colour map, note logarithmic scale) display several distinct features. The spectra are averaged in the radial direction and normalised to unit area within the shown frequency interval. The solid black line marks the value of the effective potential at the outer edge of the experimentally accessible area. Besides a few sharp peaks at $m = -9$, $-8$ and $-7$ (localised orange spots near the solid black line), the counterrotating waves manifest as broad excitations whose frequencies approximately follow the light-ring frequency (dotted black line). Conversely, the corotating waves display a set of sharp peaks (orange spots in $m>0$ bands). The white-shaded area indicates azimuthal numbers $-3 \le m \le 3$ most influenced by mechanical noise, predominantly caused by a spinning propeller that drives the vortex flow \cite{svancara2024}, and are excluded from analysis. Horizontal features between 4 and 10~Hz are also due to this noise (see the Supplementary document).}
    \label{fig:syst}
\end{figure}

%TC:ignore
\begin{figure*}[ht!]
    \centering
    \includegraphics[scale=1]{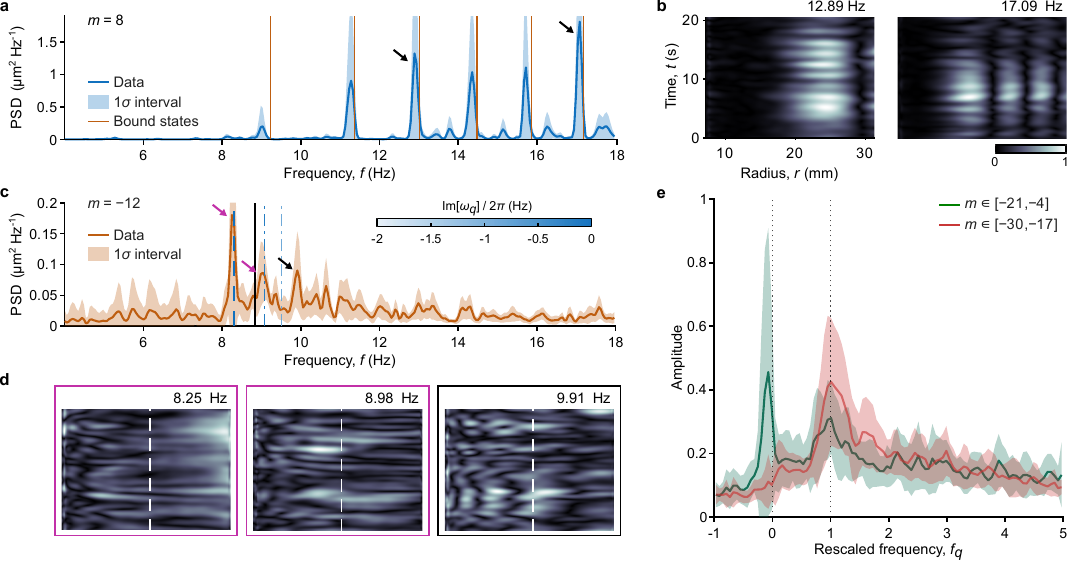}
    \caption{{\bf Ringdown spectroscopy of a confined analogue black hole.} \panel{a}~Power spectral density (PSD) estimate of corotating interface waves with azimuthal number $m=8$ averaged in the radial direction (blue line, with the blue-shaded area marking the $1\sigma$ error interval) is compared with the predicted frequencies of bound states (vertical orange lines). 
    \panel{b}~Normalised spatio-temporal amplitude diagrams (see Methods) corresponding to two bound states marked in panel~\panel{a} by black arrows.
    \panel{c}~PSD of counterrotating waves with $m = -12$ (orange line, with orange-shaded area marking the $1\sigma$ error interval) display excitations consistent with our modelling of confined quasinormal modes (QNMs, vertical blue lines). Resonances $\omega_q$ under (above) the light-ring frequency (vertical black line) are plotted as dashed (dash-dotted) lines, with their colour coding their imaginary frequency component.
    \panel{d}~Normalised spatio-temporal amplitude diagrams corresponding to resonances marked in panel~\panel{c} by purple and black arrows. Diagrams corresponding to QNMs are highlighted by purple frames. White dashed line marks the position of the light ring.
    \panel{e}~PSDs obtained by partial averaging of 27 $m$-bands in linearly rescaled frequency units $f_q$, in which the first two resonances appear at $f_q = 0$ and $1$ (dotted black lines). The averages (green and red lines with coloured areas marking the $1\sigma$ intervals, see main text) display the excitation of the first two resonances.} 
    \label{fig:spectra}
\end{figure*}
%TC:endignore

Our experimental data provide a direct means to validate these considerations. In Fig.~\ref{fig:syst}, we present the radially averaged power spectral density (PSD) of interface fluctuations across a range of azimuthal numbers $m$. We observe a distinct behaviour on the two sides of the spectrum and investigate these features separately. Corotating ($m>0$) waves exhibit distinct, localised excitations. As a representative example, Fig.~\ref{fig:spectra}a shows in detail the spectrum for $m=8$ (blue line; the blue shaded region indicates the one-standard-deviation ($1\sigma$) interval from averaging over $r$). Consistent with previous results \cite{svancara2024}, these spectral features correspond to bound states oscillating between the hard outer boundary and the effective potential barrier. Numerically determined resonance frequencies corresponding to the shape and size of this dynamically induced cavity (vertical orange lines, see Methods) closely match the experimental data. Spatio-temporal diagrams, obtained by digitally extracting the signals associated with the peaks marked by diagonal arrows (see Methods), are shown in Fig.~\ref{fig:spectra}b. These diagrams highlight the characteristic structure of bound states, with pronounced vertical dark bands indicating the nodes of these standing waves. Additional, less distinct low-frequency amplitude modulations (horizontal dark features) are also observed.

Returning to Fig.~\ref{fig:syst}, we now consider the counterrotating ($m<0$) waves. These excitations follow the light-ring frequency (dotted black line) only approximately as the modes are spread across several frequency channels. Some azimuthal bands, such as $m=-7$ and $m=-8$, feature distinct peaks below the light-ring frequency, close to the edge of our field of view (solid black line), which reflects the value of the effective potential at the boundary of the experimentally accessible region. The occurrence of frequency-shifted oscillations contrasts with earlier observations of QNMs in a dissipative hydrodynamic system \cite{torres2020}, where significant emission was observed only at the light-ring frequency, which shows that spatial confinement plays a decisive role in shaping the QNM spectrum.

The effective potential and outer boundary together form a shallow cavity whose depth and spatial extent are mainly controlled by the vortex circulation. However, such confinement arises only within a limited range of experimental parameters, particularly restricting vortex circulations. To illustrate this, Fig.~\ref{fig:intro}d shows the effective potentials at twice the circulation (grey lines), demonstrating that increasing circulation shifts the potential maxima outward and reduces the cavity depth. We accessed the regime required to observe confinement effects by systematically exploring the parameter space of interface height, circulation and temperature. Optimal experimental settings were found around $1.70$~K.

The interaction of a wave with the outlined effective potential can be treated as a scattering process with frequency-dependent reflection and transmission coefficients, $R(\omega)$ and $T(\omega)$, which determine the relative amplitudes of the reflected and transmitted waves. This approach must be combined with suitable boundary conditions. Our analysis shows that the experimental data presented below is consistent with having a full-slip (Neumann) boundary condition at the outer wall and full absorption in the vortex core. The full-slip condition captures the dynamics of superfluid waves reflecting from a solid wall, with experimental evidence of this behaviour reported in \cite{barroso2025holography}, while the fully absorbing condition models a one-directional membrane, such as the black hole horizon considered in black-hole spectroscopy.

While the QNMs of an open system appear as poles of $R(\omega)$ in the complex frequency plane, the additional constraints lead to a different set of complex resonances $\omega_q$ satisfying a resonance condition that takes into account the finite size of the experiment (see Methods). As a consequence, the QNMs of the open system migrate in the complex plane, and we can expect one or more resonances below the light-ring frequency, corresponding to quasi-bound states, and several damped overtones above it (Methods, Fig.~\ref{fig:rescond}). The damping rate associated with energy loss into the vortex core is given by the imaginary component of $\omega_q$. Modelling of our black hole analogue, building on recent theoretical insights \cite{solidoro2024}, predicts substantially reduced damping compared with the open system, where in practice only the longest-lived mode at the light-ring frequency is detectable. Conversely, spatial confinement generates multiple, weakly damped resonances with frequencies that do not necessarily coincide with the light-ring frequency. This transformative shift enables the excitation and detection of several QNMs in our system.

In Fig.~\ref{fig:spectra}c, we superimpose the predicted resonance frequencies (dashed and dot-dashed blue lines) onto the measured PSD of surface waves for $m=-12$ as an example. We observe a pronounced excitation of a quasi-bound state at $8.25~\mathrm{Hz}$, followed by a broader peak whose position agrees with the first overtone (both peaks are marked by purple arrows). The corresponding spatio-temporal diagrams (purple boxes in Fig.~\ref{fig:spectra}d) reveal a characteristic and seemingly universal pattern. The excitations extend across the full accessible radial range and thus transport energy from the outer region towards the vortex core. The ability to cross the potential barrier is a defining signature of QNMs.

The remaining diagram in Fig.~\ref{fig:spectra}d, associated with the additional peak in the $m=-12$ spectrum (Fig.~\ref{fig:spectra}c, black arrow), qualitatively differs from the previous diagrams. Resonances in a system with an open inner boundary are not expected to feature any nodal structure to the left of the light ring \cite{solidoro2024}, suggesting additional mechanisms at play (such as mode-vortex interactions) that go beyond our linear modelling and call for further explorations of the full hydrodynamical system.

Although these additional modes appear across different azimuthal bands (cf. Fig.~\ref{fig:syst} and the Supplementary document), their frequencies show no systematic pattern. This contrasts with the QNMs, which consistently emerge at the frequencies predicted by our model. To emphasise this, we use $m$ as an effective control parameter and assess how the spectrum evolves as $m$ varies. In Fig.~\ref{fig:spectra}e, we plot the PSDs (solid lines; shaded regions indicate $1\sigma$ intervals) averaged over two ranges of $m$-bands (see legend). Here, the frequency axis of each PSD is linearly rescaled to a non-dimensional variable $f_q$ such that the first two resonances map to $f_q = 0$ and $1$. The spectrum for $m\in[-21,-4]$ (green) corresponds to $m$-bands where the first two resonant modes are a quasi-bound state and an overtone. Both modes are visible in the experimental data, although the second peak is noticeably broadened due to increased dissipation of the overtone and the possible presence of additional resonances within its linewidth (cf. the overtone spacing in Fig.~\ref{fig:spectra}c). The spectrum averaged over $m\in[-30,-17]$ (red) shows only the peak at $f_q=1$. In this regime, the averaging spans $m$-bands for which this mode can lie above or below the light-ring frequency, and once it becomes a quasi-bound state, its reduced damping enhances its visibility. The absence of the $f_q=0$ peak is a technical limitation of our detection method: for the considered azimuthal numbers, the first resonance is localised between the system boundary and the edge of our field of view and its non-detection does not imply that the wave is not excited. On the contrary, weak damping of the lowest-frequency resonances makes their excitation highly likely. Together, the spectra in Fig.~\ref{fig:spectra}e demonstrate that multiple QNMs can be excited, resolved, and quantitatively understood in a controlled laboratory setting, providing a direct means for probing their dynamics.

The effects explored here address generic confinement features that extend beyond specific models considered in black-hole spectroscopy studies. Modifications of the quasinormal-mode spectrum arise in a wide range of physical and theoretical settings, including accretion disks, interstellar plasma, asymptotically AdS spacetimes, and ultralight bosonic fields~\cite{hannusekla2019,spieksma2023,Spieksma:2024voy}. The latter can develop macroscopic phase order \cite{ferreira2021} and even nucleate quantum vortices \cite{kain2010vortices,korshynska2025vortex}, bringing parallels with our superfluid system.

By leveraging the enhanced prospects for QNM detection granted by spatial confinement, our rotating black hole simulator offers a controlled setting to probe not only the fundamental quasinormal mode structure but also its elusive overtones, demonstrated here by matching predictions for the first overtone with peaks in the noise spectrum. The additional structures observed (see Fig.~\ref{fig:spectra}) indicate that substantially more information is encoded in the noise, reflecting the interplay between the full fluid dynamics of the system and the details of the driving mechanisms. Extracting meaningful physics from such data is challenging, and closely parallels current problems in gravitational-wave data analysis. There, new approaches such as simulation-based inference are being developed to extract physical information directly from complex, noise-dominated data \cite{Dax:2024mcn}. Applying these techniques to analogue gravity systems is therefore a natural next step. Recent work suggests that this is a promising route \cite{Solidoro2026}, positioning laboratory experiments as realistic test beds for next-generation data-analysis strategies developed in gravitational-wave astronomy. This synergy fosters a virtuous cycle, where insights from controlled experiments feed back into our modelling of astrophysical black holes, and vice versa.
\medskip

%TC:ignore
\paragraph*{\bf Acknowledgements}
We are grateful to Cameron R. D. Bunney, Gregorio Carullo, Christoph Eigen, Gr\'egoire Ithier, Th\'eo Torres, Maxime Jacquet, Friedrich K\"onig, Xavier Rojas, and Sebastian H. V\"{o}lkel for valuable feedback and fruitful discussions. PS, P\v{S}, SP, CFB, RG and SW extend their appreciation to the Science and Technology Facilities Council for their generous support within Quantum Simulators for Fundamental Physics (ST/T006900/1, ST/T005858/1, and ST/T00584X/1), as part of the UKRI Quantum Technologies for Fundamental Physics programme. LS and SW gratefully acknowledge the support of the Leverhulme Research Leadership Award (RL-2019-020). MR acknowledges partial support from the Conselho Nacional de Desenvolvimento Científico e Tecnológico (CNPq, Brazil, grant \mbox{315991/2023-2}), and from the São Paulo Research Foundation (FAPESP, Brazil, grants \mbox{2022/08335-0}, \mbox{2024/00923-6} and \mbox{2025/02701-3}). SW also acknowledges the Royal Society University Research Fellowship (UF120112). RG and SW acknowledge support from the Perimeter Institute. Research at Perimeter Institute is supported by the Government of Canada through the Department of Innovation, Science and Economic Development Canada and by the Province of Ontario through the Ministry of Research, Innovation and Science.

\paragraph*{\bf Author Contributions} All authors contributed substantially to the work. Acquisition and analysis of experimental data by PS, LS, and P\v{S}. Theoretical modelling developed by LS, SP, and MR. All authors contributed to the discussion of the results and the writing of the manuscript. Funding acquisition by SW, RG, and CFB. Overall conceptualization and project supervision by SW.

\paragraph*{\bf Data \& Code availability} The datasets generated and analysed during this study are available upon reasonable request. This study does not rely on custom code or algorithms beyond standard numerical evaluations.

\paragraph*{\bf Materials \& Correspondence} SW is the corresponding author. Requests for data should be addressed to P\v{S}.

\interlinepenalty=10000
\bibliography{biblio}

\clearpage\raggedbottom\newpage

\interlinepenalty=0

\section*{Methods}

\subsection{Experimental set-up}
\vspace{-1em}

The superfluid phase of helium-4 exists at temperatures lower than $2.17$~K. The experiment presented in the main text was carried out at $1.70$~K, and additional supporting experiments (see Supplementary document) were carried out between $1.62$ and $1.96$~K. We achieve sufficiently low temperatures in a glass cryostat consisting of two concentric, double-walled Dewar vessels. Thermal insulation is provided by a liquid nitrogen jacket surrounding the inner Dewar and insulation vacuum better than $10^{-7}$~mbar. The cooling power of approximately $500$~mW is provided by evaporating the liquid helium bath. A steady vortex flow within the superfluid is initiated and maintained by a centrifugal pump mechanism~\cite{yano2018}, realised by a propeller located underneath the experimental zone, and spinning at 100 rotations per minute. The propeller is switched on 60~s before acquiring experimental data, to establish a steady recirculation loop that provides mass and momentum flux to the experimental zone, and remains on during data acquisition. The experimental zone is illuminated through the cryostat's side walls, but interface fluctuations are recorded through an optically flat window at the top flange of the helium cryostat. Within the synthetic Schlieren imaging technique \cite{moisy2009,wildeman2018}, we image a periodic pattern through the superfluid interface and analyse how surface waves deform the pattern's image. Compared to our previous work \cite{svancara2024}, here we use a finer pattern and an improved imaging system consisting of a telephoto lens (Zeiss Milvus 2/135 mm) with a teleconverter (Sigma TC-2001) and a high-speed camera (Phantom VEO-640L), leading to the increased sensitivity to interface fluctuations by approximately 30\%. We typically collect videos containing 4,096 frames at resolution of $\text{1,536}\times\text{1,536}$~pixels and rate of 200~frames per second. The resulting interface height field is equidistantly sampled in the polar coordinate system by $128$ points in the azimuthal and $256$ points in the radial direction, respectively.

Our velocity fitting procedure, described in detail in \cite{svancara2024}, consists of identifying the minimum propagating frequency of interface waves from experimental data. The obtained velocity field (Fig.~\ref{fig:intro}b) is consistent with previous studies of draining vortices \cite{andersen2003}. The best fit of the azimuthal velocity with Eq.~\eqref{eq:velocity} yields a centrally confined vortex with circulation $C = (769 \pm 5)~\mathrm{mm^2\,s^{-1}}$ and a solid-body rotation frequency $\Omega = (0.10 \pm 0.02)~\mathrm{rad\,s^{-1}}$. A fundamental consequence of quantum mechanics is the quantisation of circulation in superfluid helium. The quanta take the form of quantum vortices \cite{barenghi2023}, one-dimensional topological defects each carrying a circulation quantum $\kappa = 2\pi\hbar/M = 9.97 \times 10^{-8}~\mathrm{m^2\,s^{-1}}$, where $M$ is the mass of the $^\text{4}$He atom. Therefore, the core of the giant vortex beneath the superfluid interface is a dense array of $N_C = 2\pi C/\kappa = \text{48,500}\pm 200$ quantum vortices. Since $N_C \gg 1$, the discrete nature of the vortices can be neglected, with the giant vortex core acting as a region of continuous vorticity.

To support the formation of a centrally confined vortex cluster over a sparse vortex lattice that mimics solid-body rotation in the superfluid \cite{feynman1957}, we implement a custom 3D-printed flow conditioner within the recirculation loop \cite{svancara2024}. The residual, propeller-induced rotation $\Omega$ leads to a vortex array with areal density (within the region outside the vortex core) $n_\Omega = 2\Omega/\kappa = (200\pm 40)$~cm$^{-2}$. As we discuss in the section below, this additional vorticity in the experimental zone has a negligible effect on the observed wave dynamics.

\subsection{Effective Hamiltonian formalism}
\vspace{-1em}

The dispersion relation, Eq.~\eqref{eq:disp}, describes the dynamics of waves propagating on the background of an irrotational flow, $\vect{v} = \grad\phi$. The waves are therefore modelled as plane-wave solutions for the fluctuations of the velocity potential $\phi$. Although the employed framework does not allow any vorticity degrees of freedom in these perturbations \cite{PerezBergliaffa:2001nd,patrick2018,Oliveira:2024quw}, we consider the effect of subdominant vorticity by an additional term $m\Omega$ in Eq.~\eqref{eq:branch}. Treating this term as subdominant is justified on the following grounds. The mean separation between quantum vortices responsible for the solid-body rotation of the superfluid is on the order of $\delta = \sqrt{1/n_\Omega} \approx 0.1$~mm. From the perspective of a single quantum vortex, whose core has a radius $r_c \approx 10^{-7}$~mm \cite{barenghi2023}, the waves in our set-up can be viewed as plane waves. On the scale of one wavelength $\lambda = 2\pi/||\vect{k}||$, the vortex-induced flow field modifies the wavefront by a phase shift
\begin{align}
      \Delta\Phi \approx N_\Omega \frac{||\vect{k}||r_c c}{v_g}\,,
\end{align}
where $N_\Omega = (\lambda/\delta)^2 \approx 10^{2}$ is the number of free vortices, $c$ is the speed of sound in superfluid helium and $v_g$ is the magnitude of the group velocity. For our set-up, we estimate $c/v_g \approx 10^3$ and $r_c/\lambda\approx 10^{-7}$, and hence $\Delta\Phi \approx 10^{-2}$~rad, indicating that the plane wave remains largely unaltered by these few vortices and their collective effect can be treated as subdominant compared to the irrotational flow induced by the giant quantum vortex.

In Eq.~\eqref{eq:branch}, $F(\vect{k})$ denotes the dispersion function, which reads for gravity-capillary waves,
\begin{align}
    \label{eq:dispfun}
    F(\vect{k}) = (gk + \gamma k^3)\tanh(h_0k)\,,
\end{align}
where
$k \equiv ||\vect{k}||$, $g$ is the gravitational acceleration, $\gamma = 2.22\times 10^{-6}$~m$^3$\,s$^{-2}$ is the ratio of surface tension and density \cite{donnelly1998}, evaluated for temperature $(1.70\pm 0.01)$~K at which our experiment took place, and $h_0 = 34$~mm is the equilibrium interface height.

The waves propagate in the laboratory frame with group velocity $\vect{v}_g = \grad_{\vect{k}}\omega = \vect{v} \pm \grad_{\vect{k}}\sqrt{F(\vect{k})}$. Since the group velocity varies with $k$, the waves are, in general, dispersive. Light rings can be defined in dispersive systems by introducing an effective Hamiltonian \cite{torres2018,patrick2020,patrick2020quasinormal},
\begin{align}
    \mathcal{H} = -\frac{1}{2}(\omega-\omega_D^+)(\omega-\omega_D^-)\,, \label{eq:Hamiltonian}
\end{align}
where $\omega_D^\pm$ denotes the two branches of the dispersion relation given by Eq.~\eqref{eq:branch}. In the effective Hamiltonian formalism, $\mathcal{H}$ results in a set of Hamilton's equations which can be solved for the phase space trajectories of high frequency wave fronts $(x^\mu(\lambda),k_\mu(\lambda))$, where $x^\mu=(t,\vect{x})$, $k_\mu = (-\omega,\vect{k})$ and $\lambda$ parametrises the trajectories. Since $\mathcal{H}$ is only an effective Hamiltonian, it does not have the units of energy, and the true wave energy is actually proportional to $\omega$. The condition
\begin{align}
    \mathcal{H}=0\,,
    \label{eq:hzero}
\end{align}
enforces the dispersion relation, given by Eq.~\eqref{eq:disp}, while the only non-trivial components of the wave dynamics in our stationary, axisymmetric setting follows from the radial Hamilton's equations,
\begin{align}
    \dot{r} = \partial_p\mathcal{H}\,,
    \quad\text{and}\quad
    \dot{p} = -\partial_r\mathcal{H}\,
\end{align}
where overdot denotes derivative with respect to $\lambda$. In the main text, we refer to
\begin{multline}
    \omega_D^+(p=0,m,r) = \frac{mC}{r^2} + m\Omega +\\
     + \left[\left(\frac{gm}{r} + \frac{\gamma m^3}{r^3}\right)\tanh\left(\frac{h_0 m}{r}\right)\right]^{1/2}\,,
\end{multline}
as the potential barrier for waves in our system. This is true in the following sense. At low $\vect{k}$, we can write Eq.~\eqref{eq:Hamiltonian} in the form $\mathcal{H} = -\frac{1}{2}g^{\mu\nu}k_\mu k_\nu$, which is the Hamiltonian governing geodesics of a massless particle. The connection to wave dynamics is that, in the geometric optics limit, wavefronts are traced out by geodesics of the underlying metric. In the \mbox{low-$\vect{k}$} regime, scattering of radially in- and out-going waves is governed by a Schr\"odinger-like differential equation containing an effective scattering potential $V(r) = -(\omega-\omega^+)(\omega-\omega^-)$ \cite{patrick2024primer}, where $\omega^\pm(r) = \omega^\pm_D(p=0,m,r)$ are the locations of the dispersion relation's extrema in the $p$--direction. The turning points of the scattering potential are  $V=0$, which is formally equivalent at low $\vect{k}$ to requirement $\omega=\omega^\pm$ for the frequency, at least in the approximation that the background varies slowly. When the full Hamiltonian \eqref{eq:Hamiltonian} is considered beyond the \mbox{low-$\vect{k}$} regime, the full wave equation can no longer be put in a simple Schr\"odinger-like form, preventing us from identifying a potential barrier which scatters in and out-going waves. However, the curves $\omega^\pm(r)$ do generalise to the fully dispersive system and in this sense, we can refer to $\omega^+(r)$, which governs the scattering of $\omega>0$ waves in our system, as the effective potential.

A mode trapped at the light ring follows an unstable circular trajectory around the vortex, satisfying the dispersion relation together with the following conditions,
\begin{align}
    \label{eq:hami}
    \partial_p\mathcal{H} = 0\,, \quad\text{and}\quad \partial_r\mathcal{H} = 0\,.
\end{align}
The points that satisfy Eq.~\eqref{eq:hami} are the saddle points of the dispersion function \cite{Berti:2004ju}, generalising the notion of the potential's maximum to the dispersive regime. In fact, for a fixed frequency $\omega$ and azimuthal number $m$, the first condition in Eq.~\eqref{eq:hami} returns $p=0$, while the second gives the radius of the circular orbit, denoted $r_\mathrm{sp}$. It follows that saddle points lying on the positive branch of the dispersion relation $\omega_D^+(r,p)$, i.e.~obeying Eq.~\eqref{eq:hzero}, satisfy 
\begin{align}
    \label{eq:spdisp}
    \partial_p\omega_D^+ = 0\,,
    \quad\text{and}\quad
    \partial_r\omega_D^+ = 0\,,
\end{align}
which correspond to the local maxima highlighted in Fig.~\ref{fig:intro}d by open circles.

The WKB approximation employed in our model breaks down near the saddle points because of the divergence of the corresponding wave amplitudes. A well-known matching procedure \cite{torres2018,torres2020,patrick2020quasinormal} solves this issue. Keeping $\omega$ and $m$ fixed, the radial part of the wave in the proximity of the saddle point takes the form,
\begin{multline}
\phi(r) \propto \alpha^+(r_\mathrm{sp})\exp\left[i\int_{r_\mathrm{sp}}^r p(r^\prime)dr^\prime\right]+ \\
+\alpha^-(r_\mathrm{sp})\exp\left[-i\int_{r_\mathrm{sp}}^r p(r^\prime)dr^\prime\right]\,,
\end{multline}
where $\alpha^-$ and $\alpha^+$ are the complex amplitudes of radially ingoing and outgoing modes, respectively. The transfer matrix, relating amplitudes on both sides of the saddle point $r_\mathrm{sp}^\pm$, is defined as \cite{IyserWill87,Torres2020RS,RayTracing2014},
\begin{align}
\label{eq:matrix}
\begin{bmatrix}
\alpha^+(r_\mathrm{sp}^-) \\
\alpha^-(r_\mathrm{sp}^-)
\end{bmatrix}
 = -ie^{-\pi \varrho}
 \begin{pmatrix}
     1/{R} & -1 \\
    1 & -1/{R}^*
 \end{pmatrix}
\begin{bmatrix}
\alpha^+(r_\mathrm{sp}^+) \\
\alpha^-(r_\mathrm{sp}^+)
\end{bmatrix}\,,
\end{align}
where the complex reflection coefficient ${R}$ is given by
\begin{align}
\label{eq:refcoeff}
    {R} &= -i\frac{(\mu\varrho)^{i\varrho}e^{-\varrho(i+\pi/2)}}{\sqrt{2\pi}}\Gamma(1/2-i\varrho)\,,
\end{align}
with $\mu$ equal to $-1$ for frequencies below the light ring and $+1$ otherwise, and 
\begin{align}
    \varrho = -\frac{\mathcal{H}\,\mathrm{sign}(\partial^2_p\mathcal{H})}{\sqrt{-\mathrm{det}(d^2\mathcal{H})}}\,.
\end{align}
The term $d^2\mathcal{H}$ in the denominator of the relation above denotes the Hessian of $\mathcal{H}$, and all quantities are evaluated at $p=0$ and $r=r_\mathrm{sp}$.

\subsection{Resonance conditions}

\begin{figure*}[htbp!]
    \centering
    \includegraphics[scale=1]{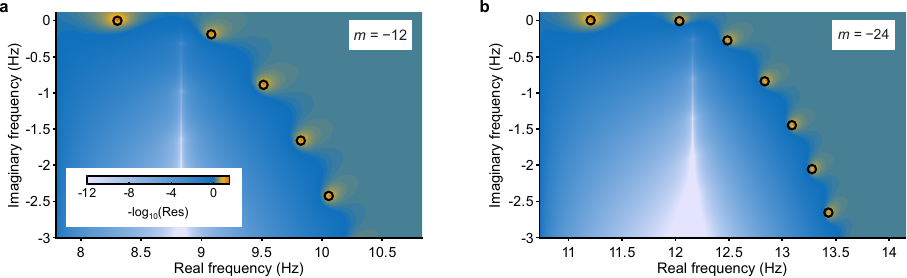}
    \caption{{\bf Identification of counterrotating resonances.} The resonance factor \eqref{eq:rescond} is evaluated as $-\log_{10}(\mathrm{Res})$ in the complex frequency plane for $m=-12$ (panel \panel{a}) and $m=-24$ (panel \panel{b}). The condition $\mathrm{Res}(\omega)=0$ is met at each local maximum, highlighted by a black circle. The bright vertical feature in each panel corresponds to the light-ring frequency. Note that we obtain one quasi-bound state, i.e. a mode oscillating below the light-ring frequency, for $m=-12$ but two for $m=-24$.}
    \label{fig:rescond}
\end{figure*}

In order to determine resonant modes, we impose a purely absorptive boundary condition at the origin by requiring $\alpha^{+}(r_\mathrm{sp}^-) = 0$. Then, the local reflection coefficient $R(\omega) = \alpha^+(r_\mathrm{sp}^+)/\alpha^-(r_\mathrm{sp}^+)$ is given by Eq.~\eqref{eq:refcoeff}. The Neumann boundary condition at the outer boundary, $r_B = 37.3$~mm, is implemented by requiring $\alpha^{+}(r_B) = \alpha^{-}(r_B)$. By relating the wave amplitude near the saddle point and the boundary, one obtains the resonance factor,
\begin{align}
    \label{eq:rescond}
    \mathrm{Res}(\omega) = R(\omega)\exp\left[2i\int_{r_-}^{r_B} p(r)dr\right] -1\,.
\end{align}
The integration limits run from the last scattering point, $r_-$, to $r_B$. For frequencies below the light-ring frequency, $r_-$ is the crossing point of a line of constant $\omega$ with the effective potential, and for higher frequencies, $r_- = r_\mathrm{sp}$.

Finally, the counterrotating resonant modes discussed in Fig.~\ref{fig:spectra}c-e must satisfy $\mathrm{Res}(\omega) = 0$ in the complex frequency plane, as shown in Fig.~\ref{fig:rescond}. The agreement between experimental data and numerical predictions confirms the validity of our model, which assumes a nearly irrotational, inviscid flow with an absorbing core.

For corotating waves, the effective potential diverges at small radii. At the crossing point with the effective potential, the amplitudes satisfy $\alpha^-(r_-) = i \alpha^+(r_-)$ \cite{patrick2020}. By combining this constraint with Neumann boundary condition at $r = r_B$, we obtain the resonance condition
\begin{align}
    \label{eq:rescond2}
    \int_{r_-}^{r_B} p(r)dr = \pi\left(n + \frac{1}{4}\right)\,,
\end{align}
where $n$ is a non-negative integer. For increasing $n$, we obtain bound states presented in Fig.~\ref{fig:spectra}a-b.

\subsection{Energy dissipation}
\vspace{-1em}

Two principal dissipation mechanisms may influence the observed excitations: viscous damping and energy absorption by the central vortex. The viscous contribution, to leading order in the kinematic viscosity $\nu$, is obtained by integrating $\nu k^{2}$ over the region where the wave propagates \cite{patrick2024primer}. A representative estimate can be made by taking the wavenumber near the outer boundary. From Fig.~\ref{fig:intro}, we infer $k\approx 0.5\,\mathrm{mm}^{-1}$ and, with $\nu\approx 10^{-8}\,\mathrm{m^{2}\,s^{-1}}$, obtain a decay rate of approximately $2.5\times 10^{-3}\,\mathrm{s^{-1}}$. Viscous damping therefore becomes relevant only on time scales of roughly $400\,\mathrm{s}$, about twenty times longer than our analysis window, validating its neglect in our experiment.

By contrast, the intrinsic dissipation of QNMs, set by $\mathrm{Im}[\omega_q]$, is more important. The damping of quasi-bound states is essentially zero (cf. Fig.~\ref{fig:rescond}), but overtones have imaginary frequencies of order $1\,\mathrm{Hz}$, corresponding to dissipation time scales of order $1\,\mathrm{s}$. This makes energy absorption by the vortex core the dominant damping mechanism, which must be balanced with the stochastic energy injection through mechanical noise. We see from Fig.~\ref{fig:rescond} that higher overtones display gradually stronger attenuation, which explains why these modes are not clearly resolved in our measurements.

\subsection{Spectral analysis}
\vspace{-1em}

The notion of direction in the experimental data is possible due to simultaneous space- and time-resolved acquisition of interface height. We decompose interface fluctuations $h(r,\theta,t)$ into individual modes by means of a two-dimensional discrete Fourier transform,
\begin{align}
    \label{eq:fft}
    \xi(r,m,f) = \frac{1}{N} \sum_\theta \sum_t e^{im\theta - i\omega t} h(r,\theta,t)\,,
\end{align}
where $\omega = 2\pi f$ and $N$ is a normalisation factor. Taking into account only the $f>0$ half-plane, the waves corotating with the vortex correspond to $m>0$ bands, while counterrotating waves correspond to $m<0$ bands.

The spectra in Fig.~\ref{fig:spectra}a,c are obtained by filtering specific azimuthal modes in the two dimensional Fourier space. Filtered signals are inverse-transformed and the resulting estimates of the power spectral density are computed using Welch's method. In comparison with $\xi$, these estimates reduce the noise level, at the expense of a reduced frequency resolution, here by a factor of 2.

The spatio-temporal amplitude diagrams $S(r,t)$ presented in Fig.~\ref{fig:spectra}b,d are computed from $\xi$ by selecting a single azimuthal number and employing a bandpass frequency filter $H$, i.e.,
\begin{align}
    S(r,t) = \left\vert\mathcal{F}^{-1}\left[\xi_m(r,f) \cdot H(f)\right]\right\vert\,,
    \label{eq:srt}
\end{align}
where $\mathcal{F}^{-1}$ indicates inverse Fourier transform in frequency and $\xi_m(r,f)$ represents a two-dimensional slice of $\xi$ at the azimuthal number $m$. For frequency filtering, we employ a raised-cosine bandpass filter, defined as
\begin{equation}
H(f) =
\begin{cases}
1 + \cos\left[\frac{\pi\left(f-f_0\right)}{w}\right] & \text{for } f \in \left(f_0-w, f_0+w\right), \\
0 & \text{elsewhere,}
\end{cases}
\label{eq:cos}
\end{equation}
where $w = 1$~Hz is the filter width.

Finally, we remark that the spectrum in Fig.~\ref{fig:syst} directly follows from $|\xi(r,m,f)|$ and the employed normalisation procedure, i.e. averaging along the radius and normalising each $m$-band to a unit area within the shown frequency interval.

\clearpage\raggedbottom\newpage

\onecolumngrid

\renewcommand\thesection{\arabic{section}}
\renewcommand\thefigure{S\arabic{figure}}    
\renewcommand\theequation{S\arabic{equation}}
\renewcommand\thetable{S\Roman{table}}

\setcounter{section}{0}
\setcounter{figure}{0}
\setcounter{equation}{0}
\setcounter{table}{0}

\begin{center}
\begin{minipage}{0.66\textwidth}
\begin{center}
        {\large \bf Supplementary document for\\Black-hole spectroscopy from a giant quantum vortex}
        \vspace{1.5em}
        
        {Pietro Smaniotto, Leonardo Solidoro, Patrik \v{S}van\v{c}ara, Sam~Patrick,\\ Maur\'icio Richartz, Carlo F. Barenghi, Ruth Gregory, and Silke Weinfurtner}
        \vspace{1.5em}
\end{center}
\end{minipage}
\begin{minipage}{0.8\textwidth}
This document provides additional evidence supporting the data presented in the main text. We start by re-analysing data presented in our previous paper \cite{svancara2024} from the viewpoint of black-hole spectroscopy. We then discuss the role of mechanical noise in our set-up and provide detailed plots of counter-rotating resonances discussed in the main text. Finally, we apply our analysis pipeline to additional data sets, revealing qualitatively similar excitation of QNMs in the superfluid simulator of curved spacetimes.
\end{minipage}
\end{center}

\section{Review of previously published data}

\begin{figure}[htbp]
    \centering
    \begin{tabular}[b]{lclc}
        \panel{a} & \itop{\includegraphics[scale=1]{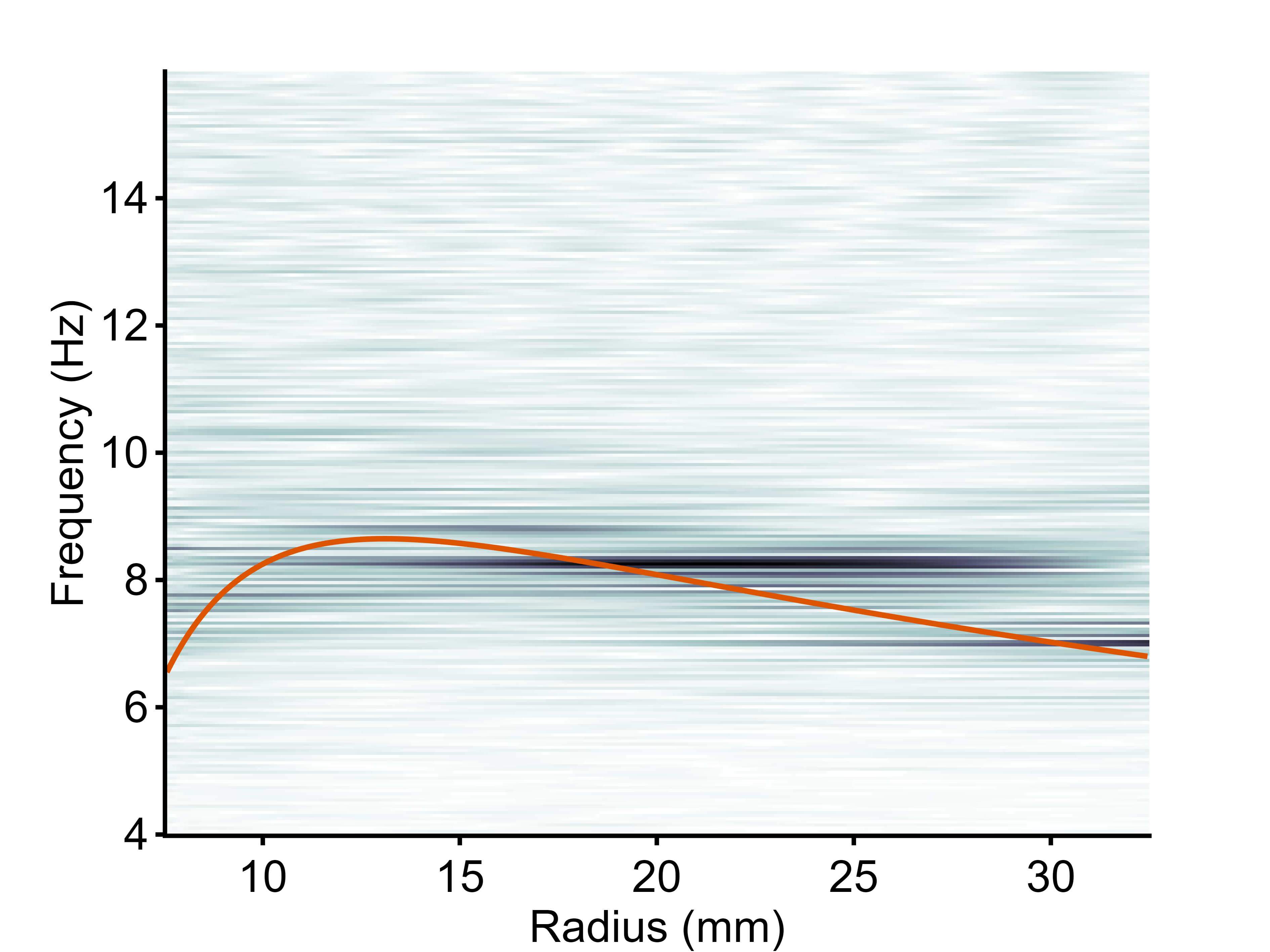}} &
        \panel{b} & \itop{\includegraphics[scale=1]{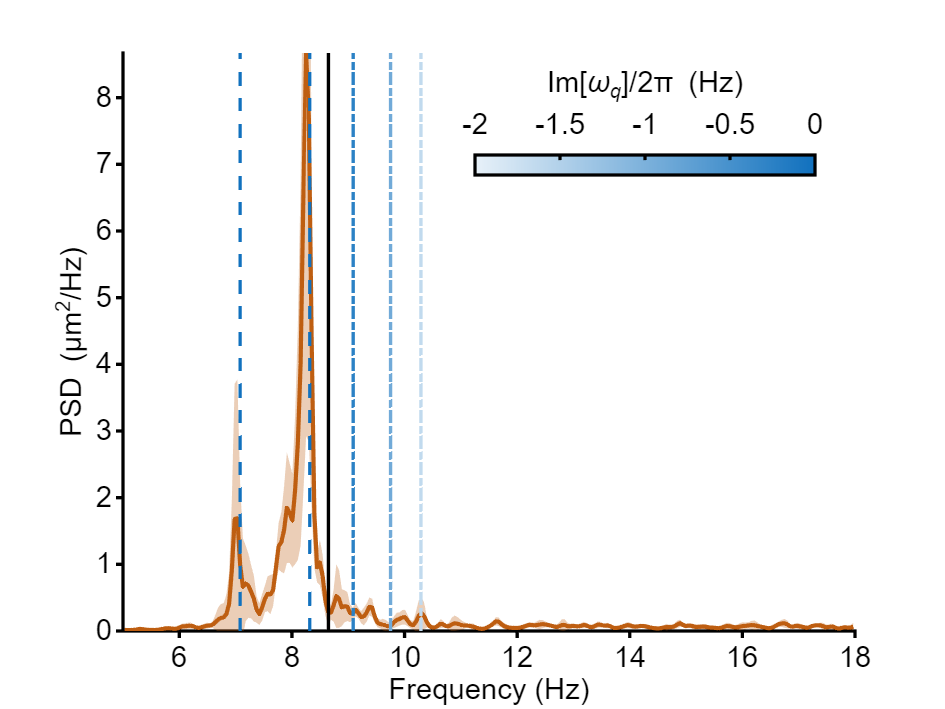}}
    \end{tabular}
    \caption{\panel{a} Spectrogram of $m = -8$ waves as a function of radius. Orange line marks the corresponding effective potential. \panel{b} Estimate of the power spectral density (orange line, with orange-shaded area indicating $1\sigma$ interval arising from radial average) is compared with the predicted counter-rotating resonances $\omega_q$ (blue lines; black line marking the light-ring frequency). Imaginary components of different $\omega_q$ are indicated by the shade of blue (see colour bar). Data corresponding to EXP~A (see Table \ref{tab:expcond}).}
    \label{fig:effpot051}
\end{figure}

Fig.~5b,c in \cite{svancara2024} presents the spectrum of waves with $m = -8$ symmetry ($m$ denoting the azimuthal number, i.e. the Fourier transformed azimuthal coordinate $\theta$). Therein, we observe that several modes are excited near the local maximum of the effective scattering potential, and claim that \emph{``these excitations, previously identified as ringdown modes of an analogue black hole, represent the very first hints of this process taking place in a quantum fluid.''} Using the modelling approach presented in the current work, we confirm that excitations shown in Fig.~5b,c indeed match the specific counter-rotating resonances. To this end, we reprocess the raw data (from now on, we refer to this data set as EXP~A) and employ a more recent velocity fitting procedure that yields circulation $C = (495\pm 3)~\mathrm{mm^2\,s^{-1}}$ and solid-body rotation $\Omega = (0.38\pm 0.01)~\mathrm{rad\,s^{-1}}$. In Fig.~\ref{fig:effpot051}a, we plot the effective potential for $m = -8$ waves (orange line) together with the corresponding Fourier spectrum as a function of radius (colour map). Two strong excitations, visible in the spectrum near 7 and 8.3~Hz, are well below the light-ring frequency (maximum of the effective potential). In Fig.~\ref{fig:effpot051}b, we show the radially-averaged estimate of the power spectral density (PSD) using the Welch method (orange line, with orange-shaded area marking the $1\sigma$ interval) and confirm the two resonances mentioned above can be explained as modes trapped between the effective potential and the outer boundary (blue dashed lines). Higher-frequency overtones (blue dot-dashed lines) do not appear to be excited beyond the ambient noise and we see no additional excitations at higher frequencies.

Although this data set displays well-defined resonances, it fails to resolve modes oscillating above the light-ring frequency, which is a crucial feature of spatially confined QNMs. This makes a systematic exploration of QNMs in EXP~A challenging, especially when compared to the more recent data sets. As a case study, we compare EXP~A with the data supporting the main text (data set titled EXP~B). We remark that EXP~A was acquired at an elevated propeller speed (175~rpm) and temperature (1.96~K); EXP~B corresponds to the propeller spinning at 100 rpm and the temperature of 1.70~K.

\begin{figure}[htbp]
    \centering
    \begin{tabular}[b]{lclc}
        & {\bf EXP~A} && {\bf EXP~B}\\
        \panel{a} & \itop{\includegraphics[scale=1]{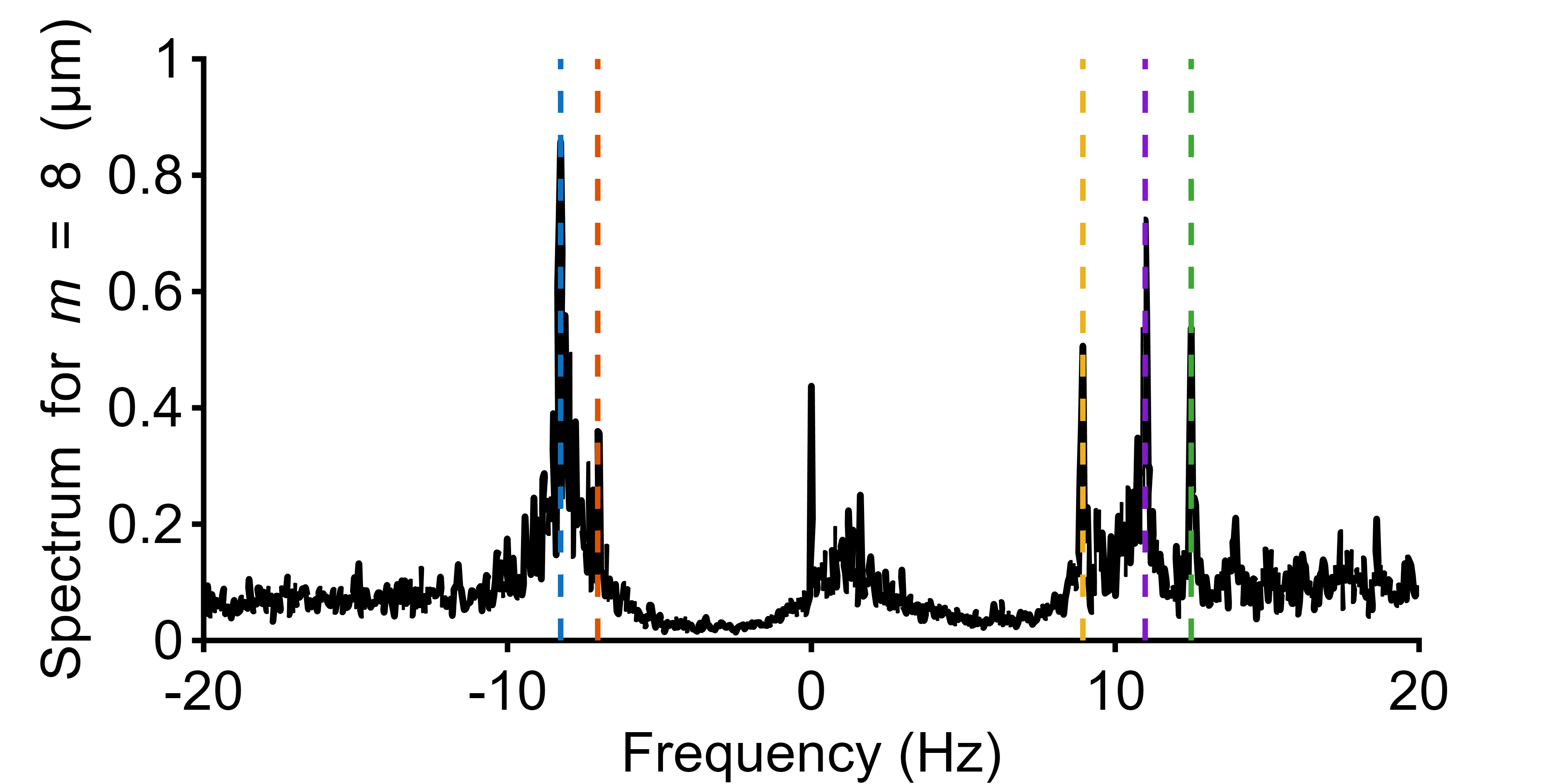}} &
        \panel{b} & \itop{\includegraphics[scale=1]{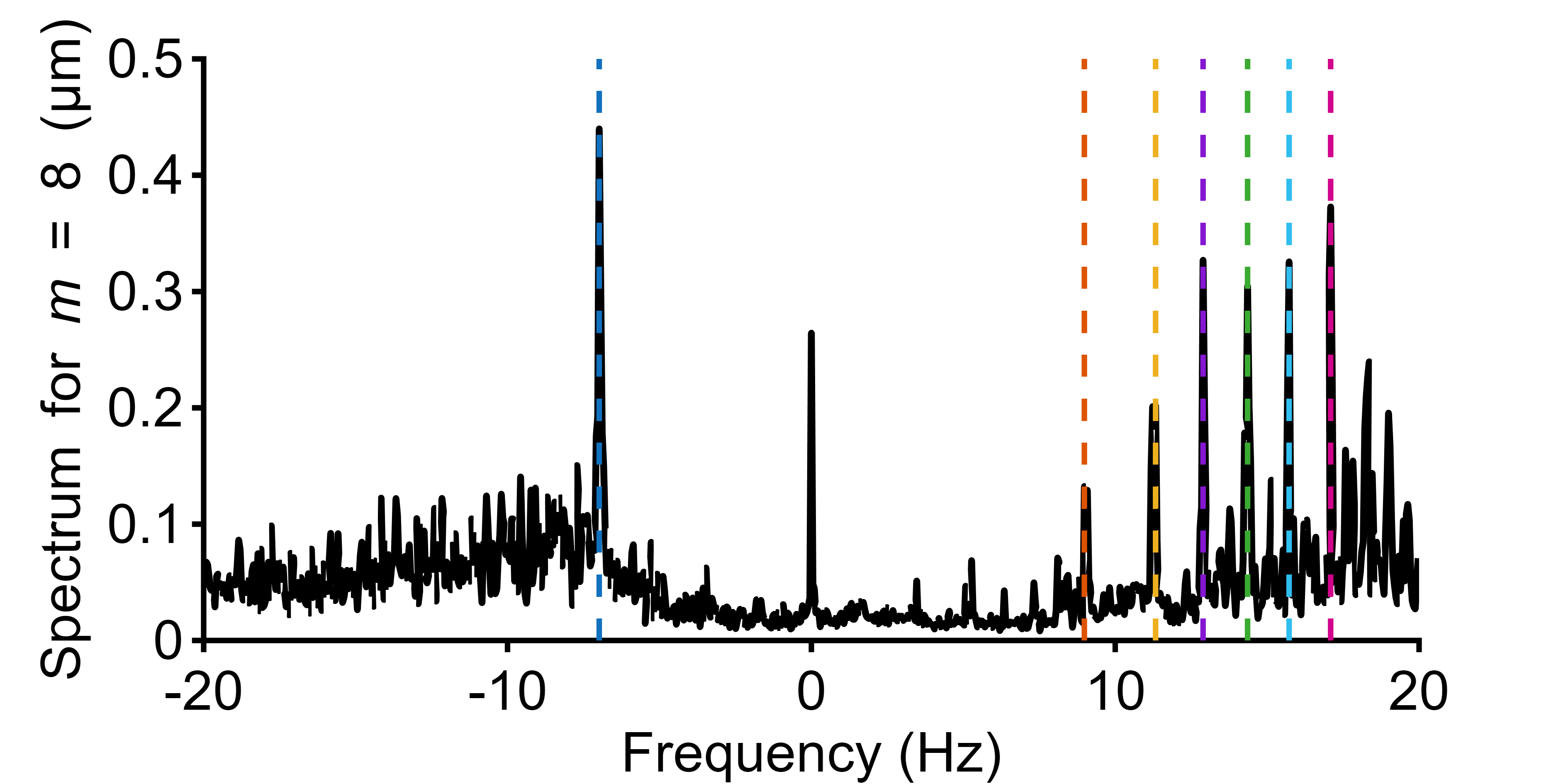}}\\
        \panel{c} & \itop{\includegraphics[scale=1]{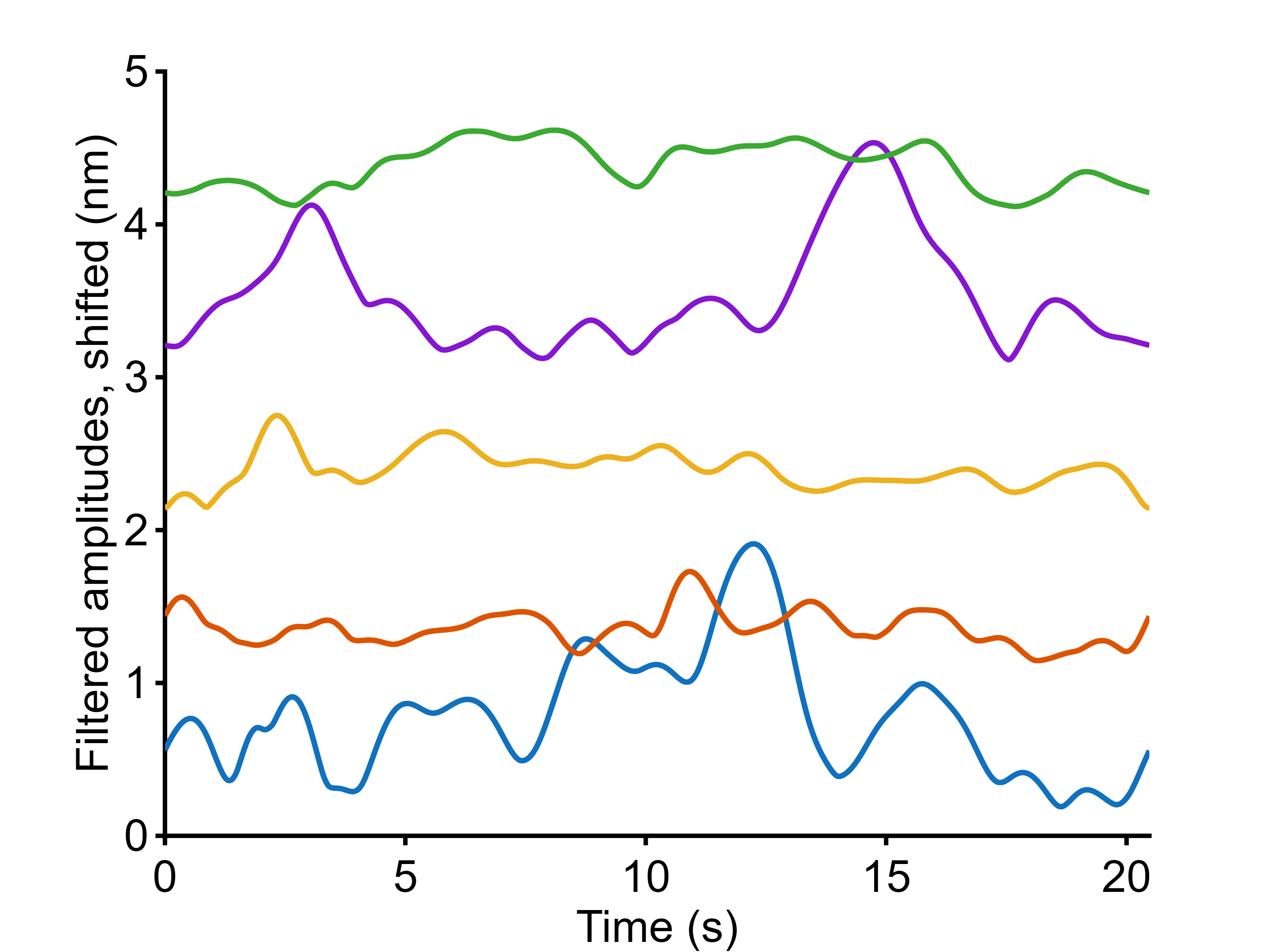}} &
        \panel{d} & \itop{\includegraphics[scale=1]{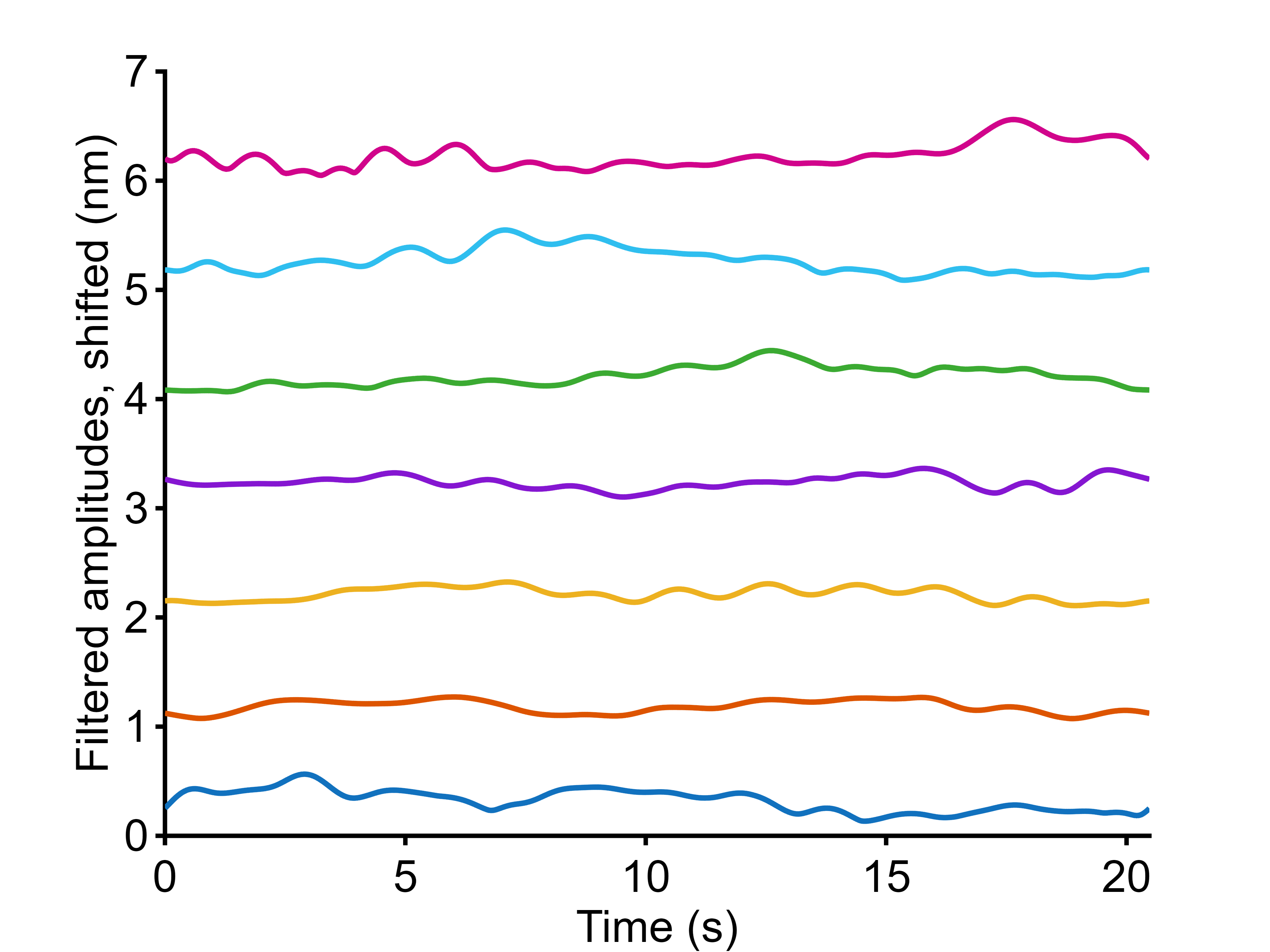}}\\
    \end{tabular}
    \caption{\panel{a,b} Frequency spectra of surface waves for $m=8$ display distinct peaks corresponding to specific modes (coloured dashed lines). \panel{c,d} Mode amplitudes \eqref{eq:at} extracted from panels a, b by a bandpass frequency filter are found to oscillate around a constant value, with more pronounced oscillations in the case of early experiments (panel c). Individual curves are vertically shifted by 1~nm for clarity.}
    \label{fig:stability}
\end{figure}

To check the stationarity of the flow/wave system, we consider waves with 8-fold azimuthal symmetry---as in \cite{svancara2024}---and analyse the time evolution of some prominent modes. These are identified in radially-averaged frequency spectra plotted in Fig.~\ref{fig:stability}a,b, and marked by coloured dashed lines. Here, modes with $f<0$ counter-rotate and those with $f>0$ co-rotate with respect to the vortex. The latter resonances correspond to bound states, standing waves trapped between the outer boundary of the experimental zone and the co-rotating effective potential that diverges at small radii. We isolate the amplitudes $A$ of these modes as a function of time by applying a bandpass frequency filter $H$ to the Fourier-transformed interface height $\xi(r,m,f) = \mathcal{F}[h(r,\theta,t)]$, performing the inverse Fourier transform, and finally averaging in radius, i.e.
\begin{equation}
    A(t) = \left\langle\left|\mathcal{F}^{-1}\left[\xi(r,m=8,f) \cdot H(f)\right]\right|\right\rangle_r\,.
    \label{eq:at}
\end{equation}
For bandpass filtering, we use a cosine filter defined as
\begin{equation}
    H(f) = 1 + \cos\left[\frac{\pi\left(f-f_0\right)}{w}\right]
    \qquad \text{for } f\in \left(f_0-w,f_0+w\right) \text{ and 0 elsewhere}\,,
    \label{eq:cos}
\end{equation}
where $f_0$ is the filter's central frequency and we choose $w = 1$~Hz.

Time evolution of mode amplitudes is shown in Fig.~\ref{fig:stability}c,d by lines whose colours correspond to central frequencies highlighted by the dashed lines in panels a and b. In both cases, the amplitudes seem to oscillate around a constant value (the curved are vertically shifted for clarity) and lack overall growth or decay. This signifies a steady vortex flow and a system of waves that reached a steady state. In case of EXP~A, the traces display more pronounced oscillations compared to EXP~B, indicating additional dynamics. We indeed notice that, in the case of EXP~A, the central depression (vortex funnel) moves inside the experimental zone, resembling the motion of an atmospheric tornado. In later experiments, including EXP~B, the motion of the funnel is suppressed. This can be attributed to the better alignment of the propeller's magnetic drive (see Methods) and a new motor that more accurately maintains a constant speed of the propeller.

\section{Characterisation of mechanical noise}

Typical design of low-temperature experiments, with equipment submerged in the liquid helium bath being suspended from the top lid of the croystat, combined with the mechanically driven flow, means that mechanical vibrations cannot be neglected. In our experiments, this noise represents and asset, and serves as a broadband source of energy. This means that stochastic vibrations of the glass cylinder surrounding the experimental zone can amplify the resonant modes, making feasible their experimental detection.

However, mechanical vibrations also induce small shifts of the backdrop pattern used for synthetic Schlieren imaging \cite{moisy2009,wildeman2018}, i.e. our detection technique. These displacements are picked by the interface reconstruction algorithm as random, low-symmetry oscillations of the superfluid interface. While most spurious signals are embedded in the $m \in \{-1, 1\}$ bands, here we exclude additional low-symmetry modes from our analysis and only consider azimuthal numbers $|m|\geq 4$.

\begin{figure}[h!]¸
    \centering
    \includegraphics[scale=1]{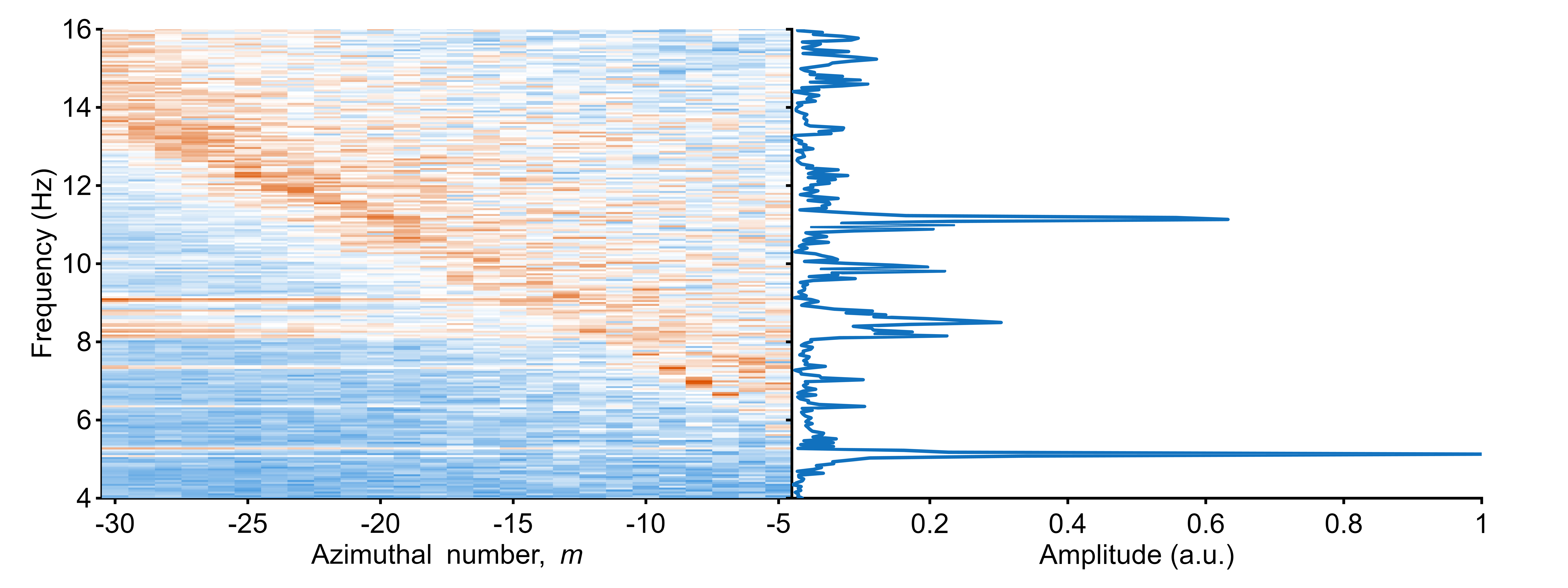}
    \caption{Frequency spectrum of the spatially-averaged interface height (right) is compared to the spectrogram presented in Fig.~\ref{fig:syst} of the main text (left). Horizontal features in the spectrogram correspond to the peaks visible in the right spectrum, reflecting the most important mechanical vibrations.}
    \label{fig:vibrations}
\end{figure}

As a first-order indication of noise in experimental data, we consider the spatially-averaged interface height corresponding to EXP~B, discussed in the main text. Since spatial averaging effectively removes all contributions from surface waves (which must be periodic along the azimuthal coordinate), what remains in the mean signal are collective oscillations due to mechanical vibrations of the set-up. In Fig.~\ref{fig:vibrations}, we reproduce part of Fig.~\ref{fig:syst} of the main text in the left panel, with the frequency spectrum of the spatially-averaged height in the right panel. Horizontal features observed in the left spectrogram coincide with spectral peaks of the spatially-averaged signal, confirming the origin of these features.

\section{Additional data sets}

The results discussed in the main text are systematically reproducible. We observe counter-rotating resonances whenever (i) the vortex flow is stationary during the period of data acquisition (typically 20~s), (ii) the level of mechanical noise present in the experiment is low but nonzero, and (iii) the flow-induced effective potential displays a shallow maximum located within the interface detection range (cf. Fig.~\ref{fig:intro}a in the main text). In some configurations, the condition (i) can be violated as the vortex may develop a sloshing instability \cite{patrick2025sloshing}. The condition (iii) is, on the other hand, given by the combined effect of temperature, frequency of the vortex drive and the free height of the superfluid. For instance, at small circulations, the effective potential develops a sharper maximum, leading to preferential excitation of waves trapped between the potential and the outer boundary, akin to co-rotating bound states.

To support our claims, we complement EXP~B presented in the main text with two additional data sets, labelled EXP~C and D. The corresponding experimental conditions, including EXP~A for the sake of completeness, are summarised in Table~\ref{tab:expcond}. EXP~C and~D were carried out at different temperatures than EXP~B and both yield a comparatively lower circulation of the central vortex.

\begin{figure}[p]
    \centering
    \includegraphics[width=\linewidth]{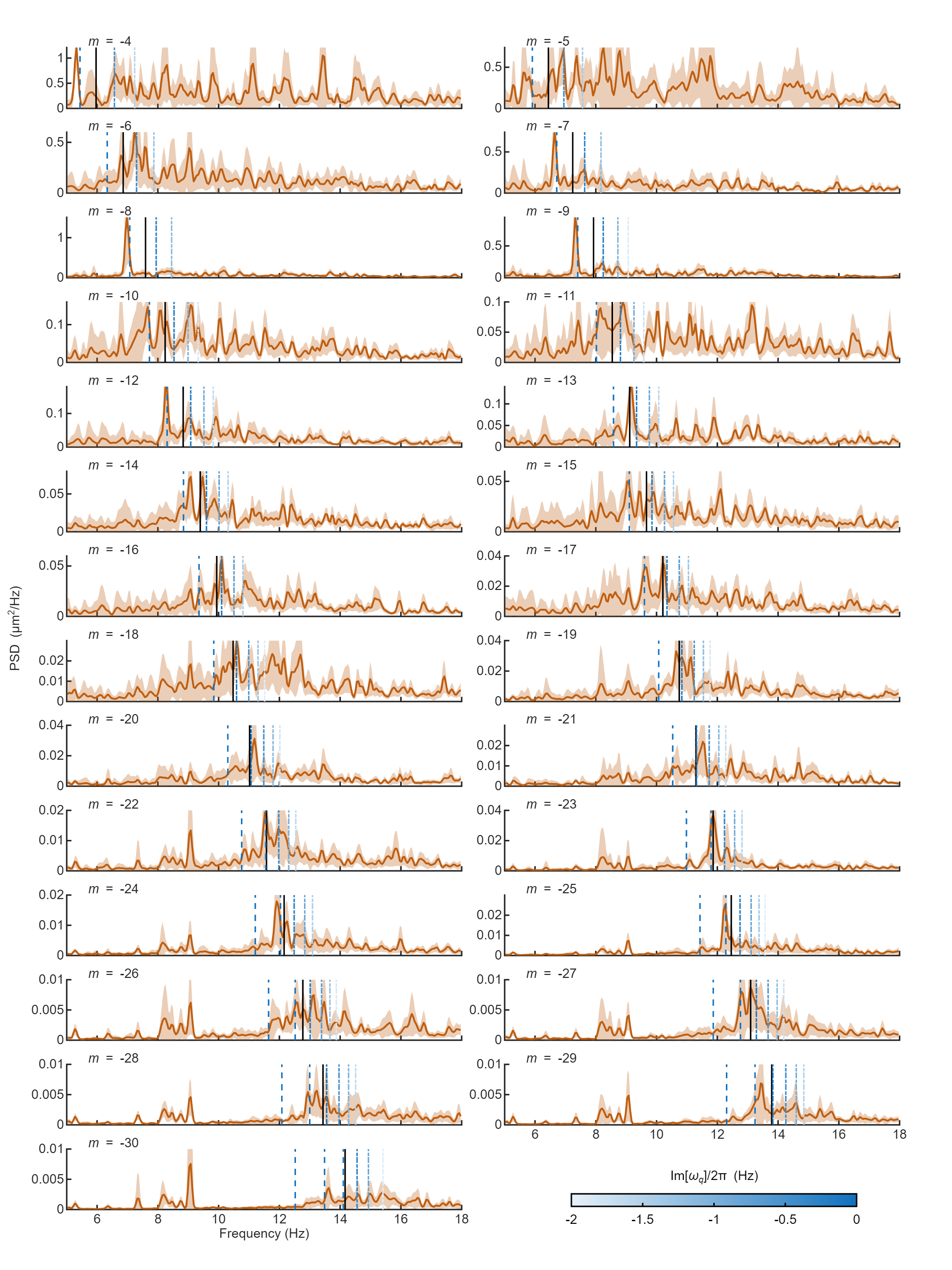}
    \caption{Power spectral density estimates for counter-rotating waves in EXP~B. Orange lines mark the radially averaged spectra; orange-shaded areas indicate $1\sigma$ statistical intervals. Blue lines denote the predicted resonances $\omega_q$: quasi-bound states (dashed lines) and overtones (dot-dashed lines). The shade of blue marks the associated imaginary frequencies (colour bar in the bottom right corner). Black line marks the corresponding light-ring frequency. }
    \label{fig:spc1}
\end{figure}

In Fig.~\ref{fig:spc1}, we show detailed frequency PSDs of interface waves from EXP~B (note that panel $m = -12$ corresponds to Fig.~\ref{fig:spectra}c in the main text). Some spectra appear to be noisier than others (e.g. $m = -6$) while other display very strong excitations well understood within our model's framework (e.g. $m = -8$ and $-9$, where the dominant excitation is identified with the lowest-frequency resonant mode). Moreover, multiple spectra display the excitation of higher-frequency resonances (e.g. $m = -12$ and $-13$) and, as discussed in the main text, these are likely linked to resonances between the vortex core and the outer boundary.

\begin{figure}[p]
    \centering
     \begin{tabular}[b]{cc}
        {\bf EXP~C} & {\bf EXP~D}\\
        \includegraphics[width=0.48\linewidth]{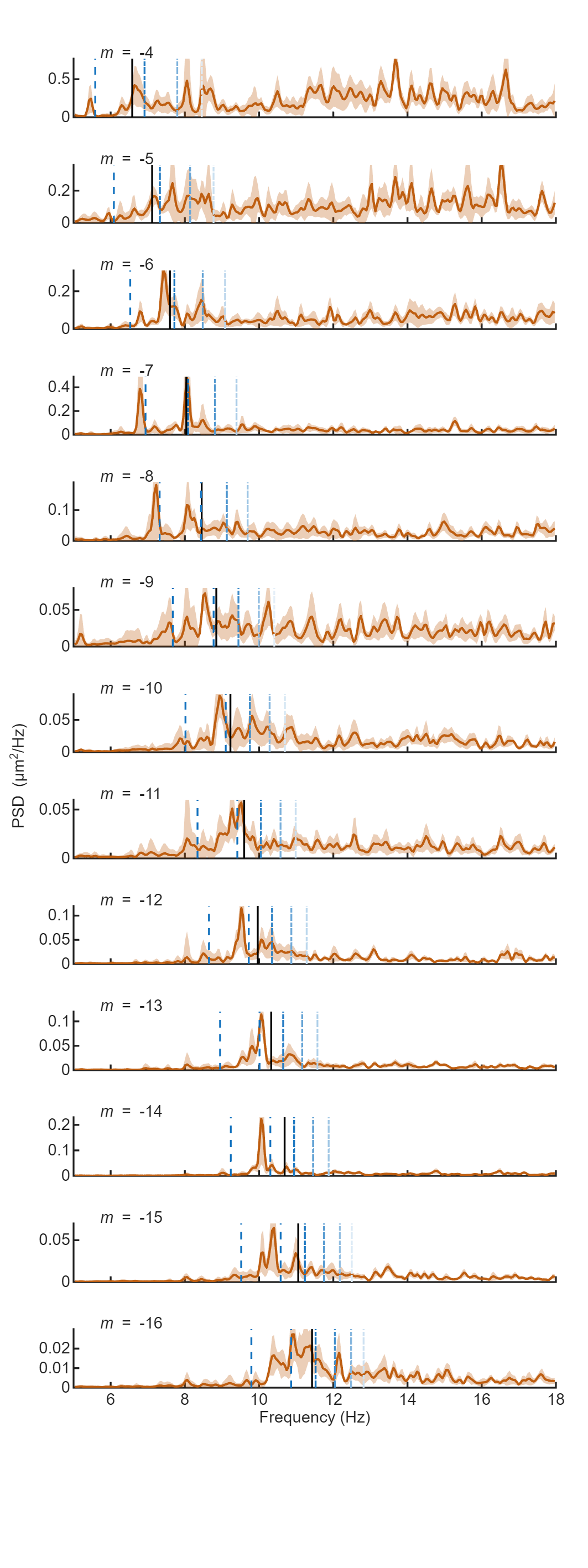} & \includegraphics[width=0.48\linewidth]{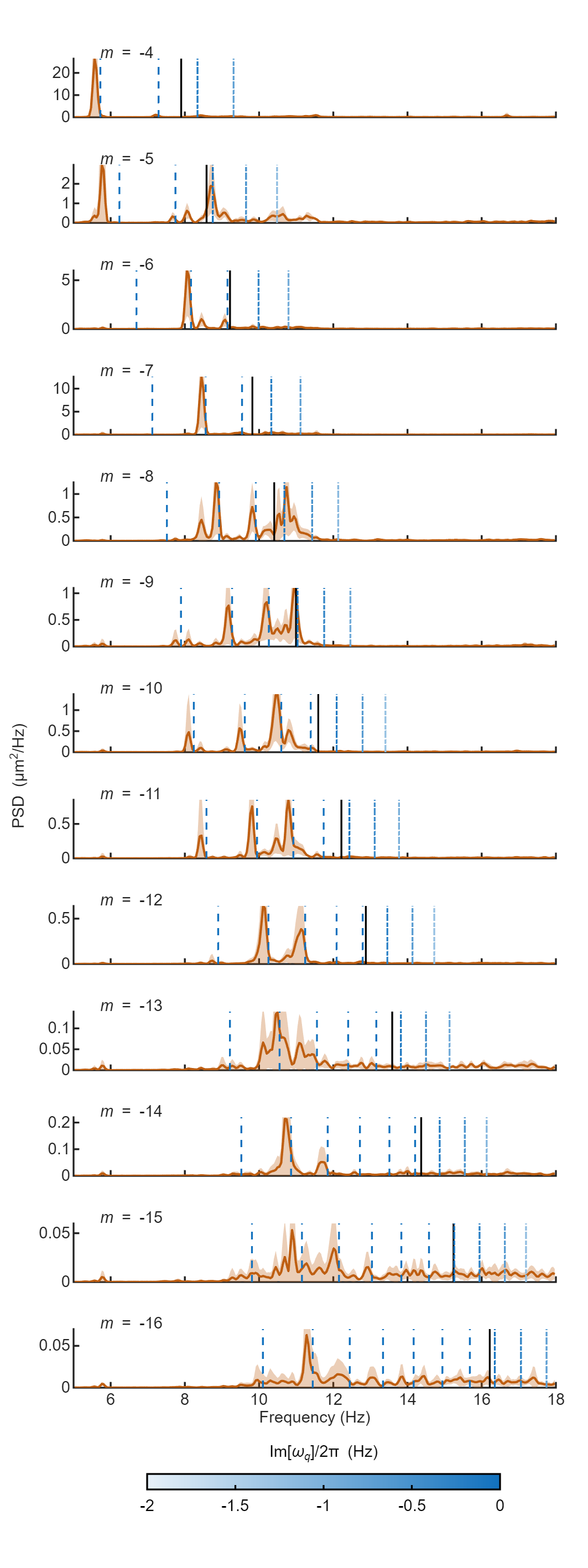}
    \end{tabular}
    \caption{Power spectral density estimates for counter-rotating waves in EXP~C (left) and EXP~D (right column) with different azimuthal numbers. Orange lines mark the radially averaged spectra; orange-shaded areas indicate $1\sigma$ statistical intervals. Blue lines denote the predicted resonances $\omega_q$: quasi-bound states (dashed lines) and overtones (dot-dashed lines). The shade of blue marks the associated imaginary frequencies (colour bar in the bottom right corner). Black line marks the corresponding light-ring frequency.}
    \label{fig:spc2}
\end{figure}

Fig.~\ref{fig:spc2} reports on similar behaviour of EXP~C and~D. In these cases, the cavity formed between the outer boundary and the effective potential allows for multiple resonances to form (blue dashed lines) and these can be recognized across both data sets and for a number of azimuthal numbers (e.g., $m = -6$ to $-12$ in EXP~D). Note that the disappearance of the lowest-frequency states is due to their localisation between the outer edge of our detection zone and the boundary. In these data sets, the identification of overtones oscillating above the maximum of the effective potential is more challenging, although we observe a few instances when these resonances are excited (e.g. $m = -7$ in EXP~C). These data sets showcase that the rich structure of QNMs is limited to some parts of the available parameter space and highlights the need for finely shaping the effective scattering potential in order to explore confinement effects affecting the QNMs in our analogue system.

\begin{table}[htb]
\caption{\label{tab:expcond}Experimental conditions for data sets EXP A-D. $T$, temperature; $h_0$, the equilibrium height of superfluid helium; $\gamma$, the ratio between surface tension and density; $f_\text{drive}$, the propeller's rotation speed; $C$, the circulation of the giant vortex; $\Omega$, the residual solid-body rotation.}
\begin{ruledtabular}
\begin{tabular}{lcccccc}
& $T$ (K) & $h_0$ (mm) & $\gamma~(\mathrm{m^3\,s^{-2}})$ & $f_\text{drive}$ (rpm) & $C~(\mathrm{mm^2\,s^{-1}})$ & $\Omega~(\mathrm{rad\,s^{-1}})$\medskip\\
EXP A (Ref.~\onlinecite{svancara2024}) & $1.96$ & $33$ & $2.10\times 10^{-6}$ & $175$ & $495 \pm 3$ & $0.38 \pm 0.01$\\
EXP B (main text) & $1.70$ & $34$ & $2.22\times 10^{-6}$ & $100$ & $769 \pm 4$ & $0.10 \pm 0.02$\\
%EXP C & $1.70$ & $28$ & $2.22\times 10^{-6}$ & $100$ & $791 \pm 4$ & $0.06 \pm 0.01$\\
EXP C & $1.62$ & $28$ & $2.25\times 10^{-6}$ & $100$ & $583 \pm 3$ & $0.10 \pm 0.01$\\
EXP D & $1.96$ & $30$ & $2.10\times 10^{-6}$ & $100$ & $346 \pm 3$ & $0.18 \pm 0.01$\\
\end{tabular}
\end{ruledtabular}
\end{table}

%TC:endignore
\end{document}